\documentstyle[11pt,epsfig]{article}
\begin{document}
\input amssym.def 
\input amssym
\hfuzz=5.0pt
%
%
%
%
\def\vec#1{\mathchoice{\mbox{\boldmath$\displaystyle\bf#1$}}
{\mbox{\boldmath$\textstyle\bf#1$}}
{\mbox{\boldmath$\scriptstyle\bf#1$}}
{\mbox{\boldmath$\scriptscriptstyle\bf#1$}}}
\def\mbf#1{{\mathchoice {\hbox{$\rm\textstyle #1$}}
{\hbox{$\rm\textstyle #1$}} {\hbox{$\rm\scriptstyle #1$}}
{\hbox{$\rm\scriptscriptstyle #1$}}}}
\def\operatorname#1{{\mathchoice{\rm #1}{\rm #1}{\rm #1}{\rm #1}}}
\chardef\ii="10
\def\widehat{\mathaccent"0362 }
\def\widetilde{\mathaccent"0365 }
\def\vphi{\varphi}
\def\vrho{\varrho}
\def\vtheta{\vartheta}
\def\ih{{\i\over\hbar}}
\def\CD{{\cal D}}
\def\CL{{\cal L}}
\def\CP{{\cal P}}
\def\CV{{\cal V}}
\def\half{{1\over2}}
\def\bhalf{\hbox{$\half$}}
\def\viert{{1\over4}}
\def\bviert{\hbox{$\viert$}}
\def\hhbox#1#2{\hbox{$\frac{#1}{#2}$}}
\def\dfrac#1#2{\frac{\displaystyle #1}{\displaystyle #2}}
\def\intT{\ih\int_0^\infty\d\,T\,e^{\i ET/\hbar}}
\def\pathint#1{\int\limits_{#1(t')=#1'}^{#1(t'')=#1''}\CD #1(t)}
\def\hbarm{{\dfrac{\hbar^2}{2m}}}
\def\hbarmq{{\dfrac{\hbar^2}{2mq}}}
\def\mzwei{\dfrac{m}{2}}
\def\overh{\dfrac1\hbar}
\def\intt{\int_{t'}^{t''}}
\def\tn{\tilde n}
\def\pmb#1{\setbox0=\hbox{#1}
    \kern-.025em\copy0\kern-\wd0
    \kern.05em\copy0\kern-\wd0
    \kern-.025em\raise.0433em\box0}
\def\bbbr{{\rm I\!R}}                                
\def\bbbn{{\rm I\!N}}                                
\def\bbbz{{\mathchoice {\hbox{$\sf\textstyle Z\kern-0.4em Z$}}
{\hbox{$\sf\textstyle Z\kern-0.4em Z$}}
{\hbox{$\sf\scriptstyle Z\kern-0.3em Z$}}
{\hbox{$\sf\scriptscriptstyle Z\kern-0.2em Z$}}}}    
\def\Cl{\operatorname{Cl}} 
\def\SU{\operatorname{SU}} 
\def\dt{\d t}
\def\d{\operatorname{d}}
\def\e{\operatorname{e}}
\def\i{\operatorname{i}}
\def\max{\operatorname{max}}
 
\begin{titlepage}
\centerline{\normalsize DESY 99--011 \hfill ISSN 0418 - 9833}
\centerline{\normalsize quant-ph/9902017 \hfill}
\centerline{February 1999\hfill}
\vskip.3in
\message{TITLE:}
\begin{center}
{\Large PATH INTEGRAL SOLUTIONS
\vskip.05in
FOR DEFORMED P\"OSCHL--TELLER-LIKE 
\vskip.05in
AND CONDITIONALLY SOLVABLE POTENTIALS}
\end{center}
\message{PATH INTEGRAL SOLUTIONS FOR DEFORMED P"OSCHL--TELLER-LIKE
AND CONDITIONALLY SOLVABLE POTENTIALS}
\vskip.5in
\begin{center}
{\Large Christian Grosche}
\vskip.2in
{\normalsize\em II.\,Institut f\"ur Theoretische Physik}
\vskip.05in
{\normalsize\em Universit\"at Hamburg, Luruper Chaussee 149}
\vskip.05in
{\normalsize\em 22761 Hamburg, Germany}
\end{center}
\normalsize
\vfill
\begin{center}
{ABSTRACT}
\end{center}
I discuss in this paper the behaviour of the solutions of the so-called
$q$-hyperbolic potentials, i.e.\ P\"oschl--Teller-like and conditionally 
solvable potentials, in terms of the path integral formalism. The differences 
in comparison to the usual P\"oschl--Teller-like potentials are investigated, 
including the discrete energy spectra and the bound state wave-functions.
\end{titlepage}
 

 
\setcounter{equation}{0}
\section{Introduction.}
\message{Introduction.}
In this paper I want to discuss some specific generalisations of 
P\"oschl--Teller related potentials. They are based on a $q$-deformation of 
the usual hyperbolic potentials, and are denoted by (we assume without loss 
of generality $q>0$)
\begin{equation}
\sinh_qx=\half(\e^x-q\e^{-x})\enspace,\qquad \cosh_qx=\half(\e^x+q\e^{-x})
\enspace.
\end{equation}
Consequently we define
\begin{equation}
\tanh_qx=\dfrac{\sinh_qx}{\cosh_qx}\enspace,\qquad 
\tanh_qx=\dfrac{\cosh_qx} {\sinh_qx}\enspace.
\end{equation}
Note the relation $\cosh_q^2x-\sinh_q^2x=q$, which has the consequence that
almost all relations known from the usual hyperbolic functions must be modified.
In analogy to the usual hyperbolic functions we have on the one hand
\begin{equation}
 \dfrac{\d}{\d x}\cosh_qx=\sinh_q x\enspace,\qquad
  \dfrac{\d}{\d x}\sinh_qx=\cosh_q x\enspace.
\end{equation}
However, on the other hand we obtain
\begin{equation}
 \dfrac{\d}{\d x}\tanh_qx=\dfrac{q}{\cosh_q^2x}\enspace,\qquad
  \dfrac{\d}{\d x}\coth_qx=-\dfrac{q}{\sinh_q^2x}\enspace.
\end{equation}
These potentials also belong the class of shape invariant potentials as 
derived from supersymmetric quantum mechanics \cite{JUNe,GENKR}, and have 
been first introduced by Arai \cite{ARAI}, and furthermore discussed by 
L\'evai \cite{LEVAI} and Lemieux and Bose \cite{LEBO} in the context of 
general solutions of the hypergeometric equation. Recently, these potentials 
have been also discussed by E\v grifes et al.\ \cite{EGRIF}. The introduction 
of the parameter $q$ may serve as an additional parameter in describing
inter-atomic interactions. The usual potentials based on the P\"oschl--Teller
type provide only at most two parameters. We can therefore investigate whether
it is possible to introduce the additional parameter $q$ to modify a sample
of known potentials which are related to the (modified) P\"oschl--Teller 
potential in order to change the energy-level feature of the potentials.

The simplest system of such a deformed potential with bound state solutions 
based on the $q$-deformed hyperbolic functions has the form ($x\in\bbbr$)
\begin{equation}
 V_1(x)=-\hbarm\dfrac{\nu^2-1/4}{\cosh_q^2x}
     =-\hbarm\dfrac{\nu^2-1/4}{\bhalf(\e^x+q\e^{-x})^2}\enspace.
\end{equation}
Extracting a factor $\sqrt{q}$ we obtain
\begin{equation}
 V_1(x)=-\dfrac{\hbar^2}{2mq}
  \dfrac{\nu^2-1/4}{\bhalf(\e^{-\ln\sqrt{q}+x}+\e^{\ln\sqrt{q}-x})^2}\enspace.
\end{equation}
If we define $y=x-\ln\sqrt{q}\in\bbbr$ we obtain
\begin{equation}
 V_1(y)=\dfrac{V_1(x)|_{q=1}}{q}=-\dfrac{\hbar^2}{2mq}\dfrac{\nu^2-1/4}{\cosh^2y}
\enspace,
\end{equation}
and the only effect is a scaling of the potential. The corresponding Lagrangian
is changed in the following way
\begin{equation}
   \CL=\dfrac{m}{2}\dot x^2+\hbarm\dfrac{\nu^2-1/4}{\cosh_q^2x}
   \quad\to\quad
   \dfrac{m}{2}\dot y^2+\dfrac{\hbar^2}{2mq}\dfrac{\nu^2-1/4}{\cosh^2y}\enspace,
\end{equation}

A more complicate version of this potential is the full modified P\"oschl--Teller
potential, now in the form of $q$-deformed hyperbolic functions, i.e.\ 
($x>\ln\sqrt{q}$)
\begin{eqnarray}
 V_2(x)&=&\hbarm\bigg(\dfrac{\eta^2-1/4}{\sinh_q^2x}-
   \dfrac{\nu^2-1/4}{\cosh_q^2x}\bigg)
   \nonumber\\   &=&
     \hbarm\bigg(\dfrac{\eta^2-1/4}{\bviert(\e^x-q\e^{-x})^2}
             -\dfrac{\nu^2-1/4}{\bviert(\e^x+q\e^{-x})^2}\bigg)\enspace.
\end{eqnarray}
Performing the transformation $y=x-\ln\sqrt{q}>0$ yields
\begin{equation}
 V_2(y)=\dfrac{V_2(x)|_{q=1}}{q}
  =\dfrac{\hbar^2}{2mq}\bigg(\dfrac{\eta^2-1/4}{\sinh^2y}
  -\dfrac{\nu^2-1/4}{\cosh^2y}\bigg)\enspace,
\end{equation}
and the only effect is again a scaling of the potential. 

For another type of these kind of potentials (Manning--Rosen potential) which is 
related to the Coulomb potential in hyperbolic geometry we define
\begin{eqnarray}
 V_3(x)&=&-\alpha\coth_qx+\dfrac{\hbar^2}{2m}
     \dfrac{\lambda^2-1/4}{\sinh_q^2x}
   \nonumber\\   &=&
  -\alpha\dfrac{\e^x+q\e^{-x}}{\e^x-q\e^{-x}}+\dfrac{\hbar^2}{2mq}
  \dfrac{\lambda^2-1/4}{(\e^{-\ln\sqrt{q}+x}-\e^{\ln\sqrt{q}-x})^2}\enspace.
\end{eqnarray}
Performing the same transformation as before we get ($x>\ln\sqrt{q}$)
\begin{equation}
 V_3(y)=-\alpha\coth y
  +\dfrac{\hbar^2}{2mq}\dfrac{\lambda^2-1/4}{\sinh^2y}\enspace,
\end{equation}
and for the Lagrangian, respectively
\begin{eqnarray}
              \CL&=&\dfrac{m}{2}\dot x^2+\alpha\coth_qx
  -\dfrac{\hbar^2}{2m}\dfrac{\lambda^2-1/4}{\sinh_q^2x}
   \nonumber\\   &\to&
   \dfrac{m}{2}\dot y^2+\alpha\coth y
  -\dfrac{\hbar^2}{2mq}\dfrac{\lambda^2-1/4}{\sinh^2y}\enspace,
\end{eqnarray}
and only the ``radial'' potential strength is now modified. Here we can expect
a significant modification of the spectral properties due to the additional
parameter $q$.

A fourth kind of potential (Rosen--Morse potential) is defined by ($x\in\bbbr$)
\begin{eqnarray}
 V_4(x)&=&\beta\tanh_qx-\dfrac{\hbar^2}{2m}\dfrac{\lambda^2-1/4}{\cosh_q^2x}
   \nonumber\\   &=&
  \beta\dfrac{\e^x-q\e^{-x}}{\e^x+q\e^{-x}}-\dfrac{\hbar^2}{2m}
  \dfrac{\lambda^2-1/4}{(\e^{-\ln\sqrt{q}+x}+\e^{\ln\sqrt{q}-x})^2}\enspace.
\end{eqnarray}
Performing the same transformation as before we get
\begin{equation}
 V_4(y)=\beta\tanh y+\dfrac{\hbar^2}{2mq}\dfrac{\lambda^2-1/4}{\cosh^2y}\enspace,
\end{equation}
and for the Lagrangian, respectively
\begin{eqnarray}
              \CL&=&\dfrac{m}{2}\dot x^2+\beta\tanh_qx
  +\dfrac{\hbar^2}{2m}\dfrac{\lambda^2-1/4}{\cosh_q^2x}
   \nonumber\\   &\to&
   \dfrac{m}{2}\dot y^2-\beta\tanh y
  +\dfrac{\hbar^2}{2mq}\dfrac{\lambda^2-1/4}{\cosh^2y}\enspace,
\end{eqnarray}
and again only the ``radial'' potential strength is modified.

Consequently, the $q$-deformed hyperbolic Scarf potential \cite{GROu} is defined by 
($x>\ln\sqrt{q}$)
\begin{eqnarray}
V_5(x)&=&V_0+V_1\coth_q^2x+V_2\dfrac{\coth_qx}{\sinh_qx}
   \nonumber\\   &\to&
   V_0+V_1\coth y+\dfrac{V_2}{\sqrt{q}}\dfrac{\coth y}{\sinh y}\enspace.
\end{eqnarray}

The $q$-deformed hyperbolic barrier potential \cite{GROu} is defined by ($x\in\bbbr$)
\begin{eqnarray}
V_6(x)&=&V_0+V_1\dfrac{\tanh_qx}{\cosh_qx}+V_2\tanh_q^2x
   \nonumber\\   &\to&
   V_0+\dfrac{V_1}{\sqrt{q}}\dfrac{\tanh y}{\cosh y} +V_2\tanh^2y\enspace.
\end{eqnarray}

There are four kinds of conditionally solvable potentials \cite{GROaf,GROag}. 
We introduce ($x\in\bbbr,r>\ln\sqrt{q}, y=x-\ln\sqrt{q},z=r-\ln\sqrt{q}$)
\begin{eqnarray}
V_7(x)           &=&\hbarm\left(-\dfrac{A\,\e^{-x}}{\sqrt{1+q\e^{-2x}}}
   +\dfrac{B}{1+q\e^{-2x}}+\dfrac{C}{(1+q\e^{-2x})^2}\right)
   \nonumber\\   &\to&
   \hbarm\left(-\dfrac{(A/\sqrt{q})\,\e^{-y/2}}{\sqrt{2\cosh y}}
   +\dfrac{B\,\e^y}{2\cosh y}+\dfrac{C\,\e^{2y}}{4\cosh^2y}\right)\enspace,
            \\   
V_7'(x)           &=&\hbarm\left(-\dfrac{A}{\sqrt{1+q\e^{-2x}}}+\dfrac{B}{1+q\e^{-2x}}
   +\dfrac{C}{(1+q\e^{-2x})^2}\right)
   \nonumber\\   &\to&
   \hbarm\left(-\dfrac{A\,\e^{y/2}}{\sqrt{2\cosh y}}\right)
   +\dfrac{B\,\e^y}{2\cosh y}+\dfrac{C\,\e^{2y}}{4\cosh^2y}\enspace,
            \\   
V_8(r)          &=&\hbarm\left(f+1-\dfrac{f-3/4}{1-q\e^{-2r}}
   +\dfrac{h_1\,\e^{-r}}{\sqrt{1-q\e^{-2r}}}+\dfrac{C}{(1-q\e^{-2r})^2}\right)
   \nonumber\\   &\to&
  \hbarm\left(f+1-\dfrac{(f-3/4)\,\e^z}{2\sinh z}
  +\dfrac{(h/\sqrt{q})\,e^{-z/2}}{\sqrt{2\sinh z}}
  +\dfrac{C\,\e^{2z}}{4\sinh^2z}\right)\enspace,
            \\   
V_8'(r)           &=&\hbarm\left(f+1-\dfrac{f-3/4}{1-q\e^{-2r}}
   +\dfrac{h_1}{\sqrt{1-q\e^{-2r}}}+\dfrac{C}{(1-q\e^{-2r})^2}\right)
   \nonumber\\   &\to&
  \hbarm\left(f+1-\dfrac{(f-3/4)\,\e^z}{2\sinh z}
  +\dfrac{h_1\,e^{z/2}}{\sqrt{2\sinh z}}
  +\dfrac{C\,\e^{2z}}{4\sinh^2z}\right)\enspace.
\end{eqnarray}
Here I have adopted the notation from \cite{GROaf,GROag}. The potentials $V_7,V_7'$
may be called ``deformed modified Rosen--Morse potential II and I'', respectively,
and the potentials $V_8,V_8'$ deformed Manning--Rosen potentials II and I'', 
respectively. These potentials are also called ``conditionally solvable'', c.f.\
\cite{DKV,GROaf,GROag,NRV} and references therein, because exact solutions can 
only be found if the parameter $C$ takes on the values $C=3/4$.
In the potentials $V_7$ and $V_8$ we indeed can change the behaviour 
of the feature of the potential, whereas in the potentials $V_7',V_8'$ there is only
a scaling due to a simple shift of coordinates.

\section{Path Integral Solution}
\subsection{The Potential $V_1$.}
We start with the potential $V_1$. According to \cite{BJB,GRSh,KLEMUS} the 
solution is given in terms of the corresponding Green function $G$ of the 
Feynman kernel. The path integral solution of the potential $V_1$ is simple,
because we can directly apply the path integral solution for the symmetric modified
P\"oschl--Teller potential \cite{KLEMUS}. Explicitly we have:
\begin{eqnarray} & &\!\!\!\!\!\!\!\!
  \pathint{x}\exp\left[\ih\intt\bigg(\mzwei\dot x^2
  +\hbarm{\lambda^2-\viert\over\cosh_q^2x}\bigg)\dt\right]\qquad
   \nonumber\\   & &\!\!\!\!\!\!\!\!
  =\pathint{y}\exp\left[\ih\intt\bigg(\mzwei\dot y^2
  +\hbarm{\tilde\lambda^2-\viert\over\cosh^2y}\bigg)\dt\right]
  =\int_{\bbbr}\dfrac{\d E}{2\pi\i}\,G^{(V_1)}(x'',x';E)\enspace,\qquad
            \\   & &\!\!\!\!\!\!\!\!
  G^{(V_1)}(x'',x';E)={m\over\hbar^2}
  \Gamma\bigg(\overh\sqrt{-2mE}-\tilde\lambda+\half\bigg)
  \Gamma\bigg(\overh\sqrt{-2mE}+\tilde\lambda+\half\bigg)
   \nonumber\\   & &\!\!\!\!\!\!\!\!\qquad\times
  P_{\tilde\lambda-1/2}^{-\sqrt{-2mE}/\hbar}(\tanh y_<)
  P_{\tilde\lambda-1/2}^{-\sqrt{-2mE}/\hbar}(-\tanh y_>)\enspace,
 \end{eqnarray}
where I have set $\tilde\lambda^2=(\lambda^2-1/4)/q+1/4$; note the relation
\begin{equation}
\tanh y=\dfrac{q^{-1/2}\e^x-q^{1/2}\,\e^{-x}}{q^{-1/2}\e^x+q^{1/2}\,\e^{-x}}
       =\tanh_qx\enspace.
\end{equation}
The $P_\nu^\mu(z)$ are Legendre functions. The bound states are given by
\begin{equation}
  \Psi^{(V_1)}(x)=\bigg(\dfrac{n-\tilde\lambda-\bhalf}{q}
  {\Gamma(2\tilde\lambda-n)\over n!}\bigg)^{1/2}
  P_{\tilde\lambda-1/2}^{n-\tilde\lambda+\half}(\tanh_qx)\enspace,
\end{equation}
and the energy spectrum is given by
\begin{equation}
E_n^{(V_1)}=-\hbarm(n-\tilde\lambda+\bhalf)^2\enspace,
\end{equation}
where $n=0,1,\dots,N_{\max}<[\tilde\lambda-\half]$, and $[x]$ denotes the integer 
values of $x\in\bbbr$.
The continuous solutions we do not state, c.f.\ \cite{GRSh,KLEMUS}.
We observe that the principal effect consists in a change in the parameter
$\lambda\to\tilde\lambda$. Depending whether $0<q<1$ or $q>1$ there is an
increasing respectively decreasing of the energy levels and the number of
energy levels in comparison to the original $1/\cosh^2x$ problem.

\subsection{The Potential $V_2$.}
Next, we consider the potential $V_2$, the deformed P\"oschl--Teller potential.
We express the solution in terms of a path integral, and again the Green function
can be stated in closed form as known from the literature 
\cite{GROu,KLEMUS}($x>\ln\sqrt{q}$)
\begin{eqnarray} & &\!\!\!\!\!\!\!\!
  \pathint{x}\exp\left\{\ih\intt\left[\mzwei\dot x^2
   -\hbarm\bigg({\eta^2-\viert\over\sinh_q^2x}
   -{\nu^2-\viert\over\cosh^2_qx}\bigg)\right]\dt\right\}
   \nonumber\\   & &\!\!\!\!\!\!\!\!
  =\!\!\!\!\pathint{y}\exp\left\{\ih\intt\left[\mzwei\dot y^2
   -\hbarmq\bigg({\eta^2-\viert\over\sinh^2y}
   -{\nu^2-\viert\over\cosh^2y}\bigg)\right]\dt\right\}
  =\int_{\bbbr}\dfrac{\d E}{2\pi\i}\,G^{(V_2)}(x'',x';E)\enspace,
   \nonumber\\   & &\!\!\!\!\!\!\!\!
            \\   & &\!\!\!\!\!\!\!\!
  G^{(V_2)}(x'',x';E)
  ={m\over\hbar^2}{\Gamma(m_1-L_\nu)\Gamma(L_\nu+m_1+1)\over
   \Gamma(m_1+m_2+1)\Gamma(m_1-m_2+1)}
   \nonumber\\   & &\!\!\!\!\!\!\!\!\qquad\times
  (q\cosh_qx'\cosh_qx'')^{-(m_1-m_2)}(\tanh_qx'\tanh_qx'')^{m_1+m_2+1/2}
   \nonumber\\   & &\!\!\!\!\!\!\!\!\qquad\times
  {_2}F_1\Big(-L_\nu+m_1,L_\nu+m_1+1;m_1-m_2+1;
                          q\cosh_q^{-2}x_<\Big)\qquad\qquad
   \nonumber\\   & &\!\!\!\!\!\!\!\!\qquad\times
  {_2}F_1\Big(-L_\nu+m_1,L_\nu+m_1+1;m_1+m_2+1;\tanh_q^2x_>\Big)\enspace,
 \end{eqnarray}
[$m_{1,2}=\half(\tilde\eta\pm\sqrt{-2mE}/\hbar)$, $L_\nu=\half(\tilde\nu-1)$,
$\tilde\eta^2=(\eta^2-1/4)/q+1/4$, $\tilde\nu^2=(\nu^2-1/4)/q+1/4$].
$_2F_1(a,b;c;z)$ is the hypergeometric function. The bound states are
\begin{eqnarray} & &\!\!\!\!\!\!\!\!
  \Psi_n^{(\eta,\nu)}(x)=N_n^{(\eta,\nu)}
         (q^{-1/2}\sinh_q r)^{\eta+1/2}(q^{-1/2}\cosh_q x)^{n-\nu+1/2}
         {_2}F_1(-n,\tilde\nu-n;1+\tilde\eta;\tanh_q^2x)\enspace,
   \nonumber\\   & &
            \\   & &\!\!\!\!\!\!\!\!
  N_n^{(\eta,\nu)}={1\over\Gamma(1+\tilde\eta)}
    \bigg[{2(\tilde\nu-\tilde\eta-2n-1)\Gamma(n+1+\tilde\eta)\Gamma(\nu-n)
    \over\Gamma(\nu-\tilde\eta-n)n!}\bigg]^{1/2}\enspace,
            \\   & &\!\!\!\!\!\!\!\!
  E_n=-\hbarm(2n+\tilde\eta-\tilde\nu-1)^2\enspace,\qquad
  n=0,1,\dots,N_{\max}<\left[\bhalf(\tilde\nu-\tilde\eta-1)\right]\enspace.
 \end{eqnarray}
Again, we omit the continuous states.

\subsection{The Potential $V_3$.}
The Manning--Rosen Potential with deformed hyperbolic functions by considering
a space-time transformation in the path integral (Duru--Kleinert transformation)
\cite{KLEo}. We have ($x>\ln\sqrt{q}$)
\begin{eqnarray} & &\!\!\!\!\!\!\!\!
\pathint{x}\exp\left[\ih\intt\bigg(\mzwei\dot x^2
      +\alpha\coth_qx-\hbarm{\lambda^2-1/4\over\sinh_q^2x}\bigg)\dt\right]
   \nonumber\\   & &\!\!\!\!\!\!\!\!
 =\pathint{y}\exp\left[\ih\intt\bigg(\mzwei\dot y^2
      +\alpha\coth y-\hbarmq{\lambda^2-1/4\over\sinh^2y}\bigg)\dt\right]
 =\int_{\bbbr}\dfrac{\d E}{2\pi\i}\,G^{(V_3)}(x'',x';E)\enspace,
   \nonumber\\   & &\!\!\!\!\!\!\!\!
            \\   & &\!\!\!\!\!\!\!\!
  G^{(V_3)}(x'',x';E)
  ={m\over\hbar^2}{\Gamma(m_1-L_E)\Gamma(L_E+m_1+1)\over
             \Gamma(m_1+m_2+1)\Gamma(m_1-m_2+1)}
   \nonumber\\   & &\!\!\!\!\!\!\!\!\qquad\times
  \Bigg({2\over\coth_qx' +1}\cdot{2\over\coth_qx''+1}\Bigg)^{(m_1+m_2+1)/2}
  \Bigg({\coth_qx' -1\over\coth_qx' +1}\cdot
        {\coth_qx''-1\over\coth_qx''+1}\Bigg)^{(m_1-m_2)/2}
   \nonumber\\   & &\!\!\!\!\!\!\!\!\qquad\times
  {_2}F_1\Bigg(-L_E+m_1,L_E+m_1+1;m_1-m_2+1;
               {\coth_qx_>-1\over\coth_qx_>+1}\Bigg)
   \nonumber\\   & &\!\!\!\!\!\!\!\!\qquad\times
  {_2}F_1\Bigg(-L_E+m_1,L_E+m_1+1;m_1+m_2+1;{2\over\coth_qx_<+1}\Bigg)\enspace,
 \end{eqnarray}
where $L_E=-\half+\sqrt{2m(\alpha-E)}/2$, and
$m_{1,2}=\half\left(2\tilde\lambda\pm\overh\sqrt{-2m(\alpha+E)}\right)$,
and $\tilde\lambda$ defined as in $V_1$.
The relevant coordinate- and time-transformations to obtain a path integral 
formulation in terms of the modified P\"oschl--Teller potential have the form
\cite{GROc,KLEMUS} (this will not be repeated here once more, $r>0$)
\begin{equation}
\half(1-\coth y)=-\dfrac{1}{\sinh^2r}\enspace,\qquad
\dt=\tanh^2r\d s\enspace.
\end{equation}
The wave functions and the energy spectrum of the bound states read $\Big(0,1,
\dots\leq N_{\max}<[\sqrt{m\alpha/2}/\hbar-\half(s+1)]$, $s=2\tilde\lambda$, 
$k_2=(1+s)/2$, $k_1=(1+(s+2n+1)/2+2m\alpha/\hbar^2(s+2n+1))/2$, note 
$n+\half-k_1<0\Big)$:
\begin{eqnarray} & &\!\!\!\!\!\!\!\!
  \Psi_n(x)=\bigg[\bigg(1+{4m|\alpha|\over\hbar(s+2n+1)^2}\bigg)
   {(2k_1-2n-s-2)n!\,\Gamma(2k_1-n-1)\over
           \Gamma(n+s+1)\Gamma(2k_1-s-n-1)}\bigg]^{1/2}
   \nonumber\\   & &\!\!\!\!\!\!\!\!\qquad\times
   \big(1-q\e^{-2x}\big)^{(s+1)/2}\e^{-(2x-\ln\sqrt{q})(k_1-s/2-n-1)}
   P_n^{(2k_1-2n-s-2,s)}(1-2q\e^{-2x})\enspace,
 \end{eqnarray}
and the energy spectrum has the form
\begin{equation}
E_n=-{\hbar^2(s+2n+1)^2\over8m}
                   -{2m\alpha^2\over\hbar^2(s+2n+1)^2}\enspace.
\end{equation}
The $P_n^{(\alpha,\beta)}$ are Jacobi polynomials.
The number of bound states is determined by $N_{\max}$, which depends on 
$\alpha$ and $s$. Decreasing $s$ for fixed $\alpha$ is archived by $0<q<1$.

\subsection{The Potential $V_4$.}
For the Rosen--Morse potential in $q$-deformed hyperbolic functions we obtain
($x\in\bbbr$)
\begin{eqnarray} & &\!\!\!\!\!\!\!\!
  \pathint{x}\exp\left[\ih\intt\!\bigg(\mzwei\dot x^2
       -\beta\tanh_qx+\hbarm{\lambda^2-1/4\over\cosh_q^2x}\bigg)\dt\right]
  \nonumber\\   & &\!\!\!\!\!\!\!\!
  =\pathint{y}\exp\left[\ih\intt\!\bigg(\mzwei\dot x^2
       -\beta\tanh y+\hbarmq{\lambda^2-1/4\over\cosh^2y}\bigg)\dt\right]
  =\int_{\bbbr}\dfrac{\d E}{2\pi\i}\,G^{(V_4)}(x'',x';E)\enspace,
   \nonumber\\   & &\!\!\!\!\!\!\!\!
            \\   & &\!\!\!\!\!\!\!\!
  G^{(V_4)}(x'',x';E)={m\over\hbar^2}{\Gamma(m_1-L_B)\Gamma(L_B+m_1+1)\over
           \Gamma(m_1+m_2+1)\Gamma(m_1-m_2+1)}
   \nonumber\\   & &\!\!\!\!\!\!\!\!\qquad\times
  \bigg({1-\tanh_qx' \over2}\cdot{1-\tanh_qx''\over2}\bigg)^{m_1-m_2\over2}
  \bigg({1+\tanh_qx' \over2}\cdot{1+\tanh_qx''\over2}\bigg)^{m_1+m_2\over2}
   \nonumber\\   & &\!\!\!\!\!\!\!\!\qquad\times
  {_2}F_1\bigg(-L_B+m_1,L_B+m_1+1;m_1+m_2+1;{1+\tanh_qx_>\over2}\bigg)
   \nonumber\\   & &\!\!\!\!\!\!\!\!\qquad\times
  {_2}F_1\bigg(-L_B+m_1,L_B+m_1+1;m_1-m_2+1;{1-\tanh_qx_<\over2}\bigg)\enspace,
\end{eqnarray}
$L_B=-\half+2\tilde\lambda$, $m_{1,2}=\sqrt{m/2}\,\big(\sqrt{-\beta-E}
\pm\sqrt{\beta-E}\big)/\hbar$. 
The relevant coordinate- and time-transformations to obtain a path integral 
formulation in terms of the modified P\"oschl--Teller potential have the form
\cite{GROc,KLEMUS} (this will not be repeated here once more, $r>0$)
\begin{equation}
\half(1+\tanh y)=\tanh^2r\enspace,\qquad
\dt=\coth^2r\d s\enspace.
\end{equation}
The wave functions and the energy spectrum 
are given by \Big($s\equiv2\tilde\lambda$; $0,\dots,n\leq N_{\max}
<[\half(s-1)-\sqrt{m|\beta|/2}/\hbar]$, $k_1=\half(1+s)$, $k_2=\half\big(1+
\half(s-2n-1)-{2mA\over\hbar(s-2n-1)}\big)>\half$\Big):
\begin{eqnarray} & &\!\!\!\!\!\!\!\!
  \Psi_n=\bigg[\bigg(1-{4m|\beta|\over\hbar(s-2n-1)^2}\bigg)
   {(s-2k_2-2n)n!\,\Gamma(s-n)\over
   \Gamma(s+1-n-2k_2)\Gamma(2k_2+n)}\bigg]^{1/2} 2^{n+(1-s)/2}
    \nonumber\\   & &\!\!\!\!\!\!\!\!\qquad\times
    (1-\tanh_qx)^{\half s-k_2-n}
    (1+\tanh_qx)^{k_2-\half}P_n^{(s-2k_2-2n,2k_2-1)}(\tanh_qx)\enspace,
            \\   & &\!\!\!\!\!\!\!\!
  E_n= -\bigg[{\hbar^2(s-2n-1)^2\over8m}
          +{2m\beta^2\over\hbar^2(s-2n-1)^2}\bigg]\enspace.
\end{eqnarray}
The number of bound states is determined by $N_{\max}$, which depends on 
$\alpha$ and $s$. Increasing $s$ for fixed $\beta$ is archived by $q>1$.

\subsection{The Potential $V_5$.}
For the $q$-deformed hyperbolic Scarf Potential we obtain ($x>\ln\sqrt{q}$,
here the coordinate transformation consists just in $x\to x/2$)
\begin{eqnarray} & &\!\!\!\!\!\!\!\!
   \pathint{x}\exp\left\{\ih\intt\!\!\left[
  \mzwei\dot x^2-\hbarm\left(V_0+V_1\coth_q^2x+V_2{\coth_qx\over\sinh_qx}
   \right)\right]\dt\right\}
   \nonumber\\   & &\!\!\!\!\!\!\!\!
  =\pathint{y}\exp\left\{\ih\intt\!\!\left[
  \mzwei\dot y^2-\hbarm\left(V_0+V_1\coth^2y
  +\dfrac{V_2}{\sqrt{q}}{\coth y\over\sinh y}
   \right)\right]\dt\right\}
 \nonumber\\   & &\!\!\!\!\!\!\!\!
  =\int_{\bbbr}\dfrac{\d E}{2\pi\i}\,G^{(V_5)}(x'',x';E)\enspace,\qquad
            \\   & &\!\!\!\!\!\!\!\!
  G^{(V_5)}(x'',x';E)={2m\over\hbar^2}{\Gamma(m_1-L_\nu)\Gamma(L_\nu+m_1+1)\over
   \Gamma(m_1+m_2+1)\Gamma(m_1-m_2+1)}
   \nonumber\\   & &\!\!\!\!\!\!\!\!\qquad\times
   \Big(q^{-1/2}\cosh_q\hhbox{x'}{2}\cosh_q\hhbox{x''}{2}\Big)^{-(m_1-m_2)}
  \Big(\tanh_q\hhbox{x'}{2}\tanh_q\hhbox{x''}{2}\Big)^{m_1+m_2+1/2}
  \nonumber\\   & &\!\!\!\!\!\!\!\!\qquad\times
  {_2}F_1\Big(-L_\nu+m_1,L_\nu+m_1+1;m_1-m_2+1;q^{1/2}\cosh_q^{-2}\hhbox{x_<}{2}\Big)
  \nonumber\\   & &\!\!\!\!\!\!\!\!\qquad\times
  {_2}F_1\Big(-L_\nu+m_1,L_\nu+m_1+1;m_1+m_2+1;\tanh_q^2\hhbox{x_>}{2}\Big)\enspace,
\end{eqnarray}
with $m_{1,2}=\eta/2\pm\sqrt{V_0+V_1-2mE/\hbar^2}$, where $\eta=
\sqrt{V_1+V_2/\sqrt{q}+1/4}$, $\nu=\sqrt{V_1-V_2/\sqrt{q}+1/4}$, 
and $L_\nu=\half(\nu-1)$. The bound-state wave-functions and the energy spectrum 
are given by
\begin{eqnarray} & &\!\!\!\!\!\!\!\!
  \Psi_n(x)=\left[{(2k_1-2k_2-2n-1)n!\,\Gamma(2k_1-n-1)\over
         2\Gamma(2k_2+n)\Gamma(2k_1-2k_2-n)}\right]^{1/2}
  \bigg(q^{-1/4}\sinh_q{x\over2}\bigg)^{2k_2-1/2}
   \nonumber\\   & &\!\!\!\!\!\!\!\!\qquad\times
  \bigg(q^{-1/4}\cosh_q{x\over2}\bigg)^{2n-2k_1+3/2}
  P_n^{[2k_2-1,2(k_1-k_2-n)-1]}\bigg({2q^{1/2}\over\cosh_q^2{x\over2}}-1\bigg)
  \enspace,\qquad\qquad
            \\   & &\!\!\!\!\!\!\!\!
  E_n=\hbarm(V_0+V_1)-\hbarm\Big[(k_1-k_2-n)-\bhalf)\Big]^2\enspace.
\end{eqnarray}
Here we denote $n=0,1,\dots,N_{\max}<k_1-k_2-1/2$, $k_1=\half(1+\sqrt{V_1-V_2/\sqrt{q}
+1/4}\,)$, $k_2=\half(1+\sqrt{V_1+V_2/\sqrt{q}+1/4}\,)$, and $\kappa=k_1-k_2-n$. In 
order that bound states can exist, it is required that $V_2<0$.

\subsection{The Potential $V_6$.}
The $q$-deformed barrier potential is treated in a similar way. 
We obtain ($x\in\bbbr$, together with the coordinate transformation $(1+\i\sinh x)/2
=\cosh^2 r$ with in order to obtain a modified P\"oschl--Teller potential in the new 
coordinate $r>0$, which will be also not be repeated again \cite{GROu})
\begin{eqnarray} & &\!\!\!\!\!\!\!\!
   \pathint{x}\exp\left\{\ih\intt\!\!\bigg[
  \mzwei\dot x^2 -\hbarm\bigg(
  V_0+V_1{\tanh_qx\over\cosh_qx}+V_2\tanh_q^2x\bigg)\bigg]\dt\right\}
   \nonumber\\   & &\!\!\!\!\!\!\!\!
  =\pathint{y}\exp\left\{\ih\intt\!\!\bigg[
  \mzwei\dot x^2 -\hbarm\bigg(
  V_0+ \dfrac{V_1}{\sqrt{q}}{\tanh y\over\cosh y}+V_2\tanh^2y\bigg)\bigg]\dt\right\}
 \nonumber\\   & &\!\!\!\!\!\!\!\!
  =\int_{\bbbr}\dfrac{\d E}{2\pi\i}\,G^{(V_6)}(x'',x';E)\enspace,\qquad
            \\   & &\!\!\!\!\!\!\!\!
  G^{(V_6)}(x'',x';E)={m\over\hbar^2}{\Gamma(m_1-L_\nu)\Gamma(L_\nu+m_1+1)\over
   \Gamma(m_1+m_2+1)\Gamma(m_1-m_2+1)}
   \nonumber\\   & &\!\!\!\!\!\!\!\!\qquad\times
  (q^{-1}\cosh_qr'\cosh_qr'')^{-(m_1-m_2)}(\tanh_qr'\tanh_qr'')^{m_1+m_2+\half}
   \nonumber\\   & &\!\!\!\!\!\!\!\!\qquad\times
  {_2}F_1\Big(-L_\nu+m_1,L_\nu+m_1+1;m_1-m_2+1;q\cosh_q^{-2}x_<\Big)
   \nonumber\\   & &\!\!\!\!\!\!\!\!\qquad\times
  {_2}F_1\Big(-L_\nu+m_1,L_\nu+m_1+1;m_1+m_2+1;\tanh_q^2x_>\Big)\enspace,
 \end{eqnarray}
with $\eta=\sqrt{V_2-\i V_1/\sqrt{q}+1/4}$, $\nu=\sqrt{V_2+\i V_1/\sqrt{q}+1/4}$, 
$L_\nu=\half(\nu-1)$, and $m_{1,2}=\eta/2\pm\sqrt{V_0+V_2/\sqrt{q}-2mE/\hbar^2}$. 
Furthermore we have $k_1=\half\sqrt{V_2/\sqrt{q}-\i V_1+\viert}\equiv\half
(1+\lambda)$, $k_2=\half(1-\lambda^*)$, with the wave-functions ($\lambda_{R,I}
=(\Re,\Im)(\lambda)$, $n=0,1,\dots,N_{\max}<[\lambda_R-\half]$)
\begin{eqnarray} & &\!\!\!\!\!\!\!\! \!\!\!\!
  \Psi_n(x)=\bigg[{(2\lambda_R-2n-1)n!\,\Gamma(\lambda-n)
   \over2\Gamma(2\lambda_R-n)\Gamma(n+1-\lambda^*)}\bigg]^{1/2}
   \nonumber\\   & &\!\!\!\!\!\!\!\! \!\!\!\!\qquad\times
  \bigg({1+\i q^{-1/2}\sinh_qx\over 2}\bigg)^{\half(\half-\lambda)}
  \bigg({1-\i q^{-1/2}\sinh_qx\over 2}\bigg)^{\half(\half-\lambda^*)}
  P_n^{(-\lambda^*,-\lambda)}(\i q^{-1/2}\sinh_qx)\enspace,\quad
 \end{eqnarray}
with the energy spectrum
\begin{equation}
E_n=\hbarm(V_0+V_2)
  -\hbarm\left\{n+\half-\sqrt{\half\left[\sqrt{\bigg(\viert+V_2\bigg)^2
      +\dfrac{V_1^2}{q}}+\viert+V_2\right]}\,\right\}^2,
\end{equation}
The energy spectrum is modified by the varying $q$ in the $V_1^2$-term.

\subsection{The Potential $V_7$.}
The solution of the path integral for the potential $V_7$ is related to the
solution of the (deformed) hyperbolic Scarf potential \cite{GROaf}. We have for
the path integral formulation
\begin{eqnarray} & &\!\!\!\!\!\!\!\!
   \pathint{x}\exp\left\{\ih\intt\!\!\left[
  \mzwei\dot x^2 - \hbarm\left(-\dfrac{A\,\e^{-x}}{\sqrt{1+q\e^{-2x}}}
   +\dfrac{B}{1+q\e^{-2x}}+\dfrac{C}{(1+q\e^{-2x})^2}\right)\right]\dt\right\}
   \nonumber\\   & &\!\!\!\!\!\!\!\!
  =\pathint{y}\exp\left\{\ih\intt\!\!\left[
  \mzwei\dot x^2 - \hbarm\left(-\dfrac{(A/\sqrt{q})\,\e^{-y/2}}{\sqrt{2\cosh y}}
   +\dfrac{B\,\e^y}{2\cosh y}+\dfrac{C\,\e^{2y}}{4\cosh^2y}\right)\right]\dt\right\}
 \nonumber\\   & &\!\!\!\!\!\!\!\!
  =\int_{\bbbr}\dfrac{\d E}{2\pi\i}\,G^{(V_7)}(x'',x';E)\enspace,\qquad
 \enspace
 \end{eqnarray}
The details of its solution are not repeated here again, c.f. also $V_5$. 
The quantization condition is found to
read ($\tilde B=B/\sqrt{q}$)
\begin{equation}
  \sqrt{A-E_n-{3\hbar^2\over8m}}
  =\half\Big(\sqrt{\tilde B-E_n}-\sqrt{-\tilde B-E_n}\,\Big)
  -{\hbar\over\sqrt{2m}}(n+\bhalf)\enspace,
\label{numf}
\end{equation}
This give after some algebra a cubic equation in $(-E_n)$
($\lambda=A+C+\tn^2,C=-3\hbar^2/8m,\tn=\hbar(n+\half)/\sqrt{2m}$)
\begin{eqnarray}       & &\!\!\!\!\!\!\!\!\!
 4 \tn^2(-E_n)^3+\Big[12\tn^2(\tn^2+\lambda)-\lambda^2\Big](-E_n)^2
         \nonumber\\   & &\!\!\!\!\!\!\!\!\!\qquad
 +\bigg[16\tn^2\lambda(A+C+\lambda)-2(\lambda+4\tn^2)
   \bigg(\lambda^2+{\tilde B^2\over4}+4\tn^2(A+C)\bigg)\bigg](-E_n)
         \nonumber\\   & &\!\!\!\!\!\!\!\!\!\qquad
  +\Bigg[16\tn^2\lambda^2(A+C)-\bigg(\lambda^2+{B^2\over4}+4\tn^2(A+C)
  \bigg)^2\Bigg]=0\enspace.
\end{eqnarray}
From the Green function of the hyperbolic Scarf-like potential
we derive the Green function for the potential $V_7$
\begin{eqnarray}       & &\!\!\!\!\!\!\!\!\!
  G^{(V_7)}(x'',x';E)=\big(\coth u'\coth u'')^{1/2}
  {2m\over\hbar^2}{\Gamma(m_1-L_\nu)\Gamma(L_\nu+m_1+1)\over
   \Gamma(m_1+m_2+1)\Gamma(m_1-m_2+1)}
         \nonumber\\   & &\!\!\!\!\!\!\!\!\!\qquad\times
 (\cosh u'\cosh u'')^{-(m_1-m_2)}(\tanh u'\tanh u'')^{m_1+m_2+\half}
         \nonumber\\   & &\!\!\!\!\!\!\!\!\!\qquad\times
  {_2}F_1\bigg(-L_\nu+m_1,L_\nu+m_1+1;m_1-m_2+1;
                          {1\over\cosh^2u_<}\bigg)
         \nonumber\\   & &\!\!\!\!\!\!\!\!\!\qquad\times
  {_2}F_1\bigg(-L_\nu+m_1,L_\nu+m_1+1;m_1+m_2+1;\tanh^2u_>\bigg)
\end{eqnarray}
with $\sinh u=\e^y=\e^{x-\ln\sqrt{q}}$, $m_{1,2}=\eta/2\pm\sqrt{V_0+V_1-8mE/\hbar^2}$,
where $\eta=\sqrt{V_1+V_2+1/4},L_\nu=\half(\nu-1)$ and $\nu=\sqrt{V_1-
V_2+1/4}$, together with the identification $V_0=2mA/\hbar
^2-\half$, $V_1=-(2mE/\hbar^2+\viert)$, $V_2=-2m\tilde B/\hbar^2$.
The poles of the Green function determine the
energy-spectrum, and the corresponding residua give the wave-functions
expansions. We obtain ($k_1=\half(1+\nu),k_2=\half(1+\eta)$, $\eta=
\sqrt{-2m(E_n+\tilde B)}/\hbar$, $\nu=\sqrt{2m(\tilde B-E_n)}/\hbar$)
\begin{eqnarray}       & &\!\!\!\!\!\!\!\!
  E_n=\sqrt[\scriptstyle3\,]{\sqrt{D}+{Q\over2}}
      -\sqrt[\scriptstyle3\,]{\sqrt{D}-{Q\over2}}+{R\over3}\enspace,
                  \\   & &\!\!\!\!\!\!\!\!
  \left.\begin{array}{l}\displaystyle
  D=\bigg({P\over3}\bigg)^3+\bigg({Q\over2}\bigg)^2\enspace,\qquad
  P={3S-R^2\over3}\enspace,\qquad Q={2R^3\over27}-{RS\over3}+T
  \enspace,       \\[3mm]\displaystyle
  R={12\tn^2(\tn^2+\lambda)-\lambda^2\over4\tn^2}\enspace,\quad
  T={16\tn^2\lambda^2(A+C)-\big[\lambda^2+\tilde B^2/4+4\tn^2(A+C)\big]^2
      \over4\tn^2}\enspace,     \\ [3mm]\displaystyle
  S={8\tn^2\lambda(A+C+\lambda)-(\lambda^2+4\tn^2)
     (\lambda^2+\tilde B^2/4+4\tn^2(A+C)\big]\over2\tn^2}\enspace.
  \end{array}\quad\right\}
 \end{eqnarray}
We omit the details concerning the wave functions. Bound states exist if 
$A<0,0<\tilde B<|A|$, and the number $N_{\max}$ of bound states is found by requiring 
$|E_n|>\tilde B$. 

\subsection{The Potential $V_8$.}
The path integral for the potential $V_8$ is related to the path integral for the
hyperbolic barrier potential as discussed in \cite{GROag}
\begin{eqnarray} & &\!\!\!\!\!\!\!\! \!\!\!\!
   \pathint{x}\exp\left\{\ih\intt\!\!\left[
  \mzwei\dot x^2 - \hbarm\left(f+1-\dfrac{f-3/4}{1-q\e^{-2r}}
   +\dfrac{h_1\,\e^{-r}}{\sqrt{1-q\e^{-2r}}}+\dfrac{C}{(1-q\e^{-2r})^2}\right)
   \right]\dt\right\}
   \nonumber\\   & &\!\!\!\!\!\!\!\! \!\!\!\!
  =\pathint{z}
  \nonumber\\   & &\!\!\!\!\!\!\!\! \!\!\!\!\quad\times
   \exp\left\{\ih\intt\!\!\left[
  \mzwei\dot x^2 - \hbarm\left(f+1-\dfrac{(f-3/4)\,\e^z}{2\sinh z}
  +\dfrac{(h/\sqrt{q})\,e^{-z/2}}{\sqrt{2\sinh z}}
  +\dfrac{C\,\e^{2z}}{4\sinh^2z}\right)\right]\dt\right\}
 \nonumber\\   & &\!\!\!\!\!\!\!\! \!\!\!\!
  =\int_{\bbbr}\dfrac{\d E}{2\pi\i}\,G^{(V_8)}(x'',x';E)\enspace,\qquad
 \enspace
 \end{eqnarray}
The details of its solution are not repeated here again, c.f.\ \cite{GROag} Due to 
the fact that its solution is defined in the half-space $\bbbr^+$, we must construct 
the corresponding Green function in terms of the Green function in the entire $\bbbr$,
a method described in \cite{GROr}. This has also been discussed in detail in
\cite{GROag} which is not repeated here. Hence we obtain
($\zeta(z)=\half(1+\tanh z), z=r-\ln\sqrt{q}>0$)
\begin{equation}
  G^{(V_8)}(E)(x'',x';E)=G(\zeta'',\zeta';E)-
  {G\big(\zeta'',\zeta(0);E\big)G\big(\zeta(0),\zeta';E)
       \over G\big(\zeta(0),\zeta(0);E\big)}
\enspace,
\end{equation}
with the Green function $G(E)$ given by
\begin{eqnarray}       & &\!\!\!\!\!\!\!\!\!\!
  G(\zeta'',\zeta';E)=\dfrac{m/\hbar^2}{\sqrt{\zeta(z')\zeta(z'')}}
  {\Gamma(m_1-L_\nu)\Gamma(L_\nu+m_1+1)\over
   \Gamma(m_1+m_2+1)\Gamma(m_1-m_2+1)}
         \nonumber\\   & &\!\!\!\!\!\!\!\!\!\!\qquad\times
  \left({1-\sqrt{\zeta(z')}\over2}\cdot
        {1-\sqrt{\zeta(z'')}\over2}\right)^{(m_1-m_2)/2}
  \left({1+\sqrt{\zeta(z')}\over2}\cdot
        {1+\sqrt{\zeta(z'')}\over2}\right)^{(m_1+m_2+1/2)/2}
         \nonumber\\   & &\!\!\!\!\!\!\!\!\!\!\qquad\times
  {_2}F_1\left(-L_\nu+M_1,L_\nu+m_1+1;m_1+m_2+1;
                 {1+\sqrt{\zeta_>(z)}\over2}\,\right)
         \nonumber\\   & &\!\!\!\!\!\!\!\!\!\!\qquad\times
  {_2}F_1\left(-L_\nu+M_1,L_\nu+m_1+1;m_1-m_2+1;
                 {1-\sqrt{\zeta_<(z)}\over2}\,\right)\enspace.
\end{eqnarray}
Here I have used the abbreviations
\begin{equation}
  L_\nu=\half\left(\sqrt{f+1+\i h_1-{2m\over\hbar^2}E}-1\right)
   \enspace,\qquad      
   m_{1,2}=-\half\sqrt{f+1-\i h_1-{2m\over\hbar^2}E}\pm
             \sqrt{\viert-f}\enspace.
\end{equation}
Note that the minus-sign in the first term in $m_{1,2}$ is due to the
reality condition of the problem \cite{GROag}. Bound states with
energy $E_n$ are determined by the equation
\begin{equation}
  {_2}F_1\Big(-L_\nu(E_n)+m_1(E_n),L_\nu(E_n)+m_1(E_n)+1;
               m_1(E_n)+m_2(E_n)+1;\bhalf\Big)\enspace.
\end{equation}
A more detailed numerical investigation of this transcendental equation involving
the hypergeometric function is left to the reader.

\section{Summary and Discussion.}
\message{Summary and Discussion.}
The results of our investigation of the introduction of the $q$-deformed
hyperbolic potentials show a combination of a shift of the coordinate
origin of the potential combined with a scaling of the potential strength.
In the cases of the potentials $V_1$ to $V_6$ the introduction of the parameter
$q$ the energy levels and the wave functions were modified by a nonlinear, however
simple way. In particular the energy levels could be easily derived from previous
calculations. The cases of the potentials $V_7$ and $V_8$ were somewhat more
difficult, which was due the fact that the energy levels are determined by a
third-order equation and a transcendental equation, respectively. $q$ also 
entered the expressions nonlinearly. Taking into account the potentials $V_7'$
and $V_8'$ we would obtain energy spectra determined by a fourth-order equation and
a transcendental equation, modified by a simple shift due to the coordinate 
translation.

Therefore these potentials can serve as modeling potentials where a finite
potential trough is required for particle interaction in molecular, atomic
or nuclear physics. This feature is in particular seen, if the potential is 
defined in the half-space $x>\ln\sqrt{q}$. Depending whether $0<q<1$ or
$q>1$ the number of energy levels and the ground state energy can be increased
or decreased, respectively. We see the convenience of the path integral 
formalism in the solutions of the deformed potential problems. We can easily 
use previous results, adapted accordingly to the present problems.
In some way, the potentials constructed from $q$-deformed hyperbolic functions 
model in a very simple and convenient way coordinate translations. In the 
``radial'' problems the introduction of the parameter $q$ forces the quantum
motion to take place in the half-space $x>\ln\sqrt{q}$ and {\it not\/} in the
half-space $x>0$. We therefore have introduced an impenetrable finite wall
between the particle motion and the coordinate origin, which may be identified
for instance with the center-of-mass location of a molecule. This feature alters
the energy spectrum in a nonlinear way, in particular if in the $q=1$-case there
is an integer quantum number $\lambda\equiv l\in\bbbn$.
However, this is a phenomenological feature and does not make new physics.

One should also keep in mind that the $q$-deformed hyperbolic potentials can be
used to describe curvature in spaces of negative constant curvature, i.e., on
hyperboloids (compare also \cite{CHb} for the interrelation of a deformed 
algebra and the constant negative curvature in the model of the hyperbolic 
plane \cite{GROf}). Let us consider the simplest hyperboloid 
\begin{equation}
  u_0^2-u_1^2-u_2^2=R^2\enspace,\qquad u_0\geq0\enspace,
\end{equation}
which describes one sheet of the double-sheeted hyperboloid $\Lambda^{(2)}$.
According to \cite{GROPOc,OLE} on $\Lambda^{(2)}$ there are nine coordinate
systems which allow separation of variables in the Helmholtz, respectively
Schr\"odinger equation. We consider the usual spherical system
($\tau\in\bbbr,\vphi\in[0,2\pi)$):
\begin{equation}
\left.\begin{array}{l}
u_0=R\cosh\tau\enspace,\\
u_1=R\sinh\tau\cos\vphi\enspace,\\
u_2=R\sinh\tau\sin\vphi\enspace,
\end{array}\right\}\qquad\longrightarrow\qquad\left\{\begin{array}{l}
u_0=\cosh_q\tau\enspace,\\
u_1=\sinh_q\tau\cos\vphi\enspace,\\
u_2=\sinh_q\tau\sin\vphi\enspace,
\end{array}\right.
\label{cosy}
\end{equation}
and we observe that with the identification $q=R^2$ the $q$-deformed spherical
coordinate system is a possible separating coordinate system for $\Lambda^{(2)}$.
Furthermore we obtain $-(\dot u_0^2-\dot u_1^2-\dot u_2^2)=q\dot\tau^2
+\sinh_q^2\tau\dot\vphi^2$. A calculation shows that the introduction of $q$ does
not change the energy spectrum features for the free quantum motion on 
$\Lambda^{(2)}$ (just rescale $m\to m/q$). We can also consider the Higgs 
oscillator $V(\vec u)=(mR^2\omega^2/2)(u_1+u_2^2)/u_0^2$ and the Coulomb 
potential $V(\vec u)=-(\alpha/R)(u_0/\sqrt{u_1^2+u_2^2}$ \cite{GROPOc}, and we 
find that the identification $R^2=q$ for the coordinate systems (\ref{cosy}) 
of all spectral properties of the two potentials remains valid. Therefore we can
interpret the deformation parameter $q$ in the hyperbolic potentials as a
{\it curvature} term.

\input cyracc.def
\font\tencyr=wncyr10
\font\tenitcyr=wncyi10
\font\tencpcyr=wncysc10
\def\cyrrm{\tencyr\cyracc}
\def\cyrit{\tenitcyr\cyracc}
\def\cyrcp{\tencpcyr\cyracc}

\bigskip\bigskip 
\vbox{\centerline{\ }
\centerline{\quad\epsfig{file=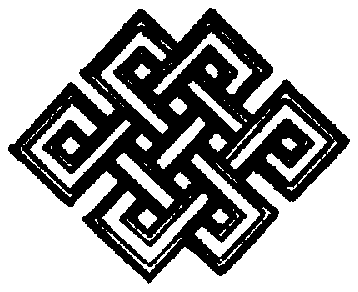,width=4cm,angle=90}}}

\end{document}